\documentclass[preprint]{aastex}

\newcommand\ubvri{\mbox{$U\!BV\!RI$}}

\begin{document}

\title{A Search for Intrinsic Polarization in O Stars with Variable Winds}

\author{David McDavid\altaffilmark{1}}
\affil{Limber Observatory, Timber Creek Road, P.O.~Box 63599,
Pipe Creek, Texas 78063-3599}
\email{mcdavid@limber.org}
\altaffiltext{1}{Guest Observer, McDonald Observatory, The University
of Texas at Austin.}

\begin{abstract}

New observations of 9 of the brightest northern O stars have been made
with the Breger polarimeter (Breger 1979) on the 0.9~m telescope at
McDonald Observatory and the AnyPol polarimeter (McDavid 1999) on the
0.4~m telescope at Limber Observatory, using the Johnson-Cousins
$\ubvri$ broadband filter system.  Comparison with earlier
measurements shows no clearly defined long-term polarization
variability.  For all 9 stars the wavelength dependence of the degree
of polarization in the optical range can be fit by a normal
interstellar polarization law.  The polarization position angles are
practically constant with wavelength and are consistent with those of
neighboring stars.  Thus the simplest conclusion is that the
polarization of all the program stars is primarily interstellar.

The O stars chosen for this study are generally known from ultraviolet
and optical spectroscopy to have substantial mass loss rates and
variable winds, as well as occasional circumstellar emission.  Their
lack of intrinsic polarization in comparison with the similar Be stars
may be explained by the dominance of radiation as a wind driving force
due to higher luminosity, which results in lower density and less
rotational flattening in the electron scattering inner envelopes where
the polarization is produced.  However, time series of polarization
measurements taken simultaneously with H$\alpha$ and UV spectroscopy
during several coordinated multiwavelength campaigns suggest two cases
of possible small-amplitude, periodic short-term polarization
variability, and therefore intrinsic polarization, which may be
correlated with the more widely recognized spectroscopic variations.

\end{abstract}


\section{Introduction}

While it is well understood that the luminosities of O stars are high
enough to cause mass loss in the form of radiation-driven winds, it is
not so easily understood that the outflows are observed to be both
episodically and also periodically or quasi-periodically variable
(Kaper \& Fullerton 1998).  By analogy with the slightly cooler Be
stars, it might be expected that the winds of the most rapidly
rotating O stars would be equatorially concentrated, even if the lower
densities and higher velocities were to prevent the formation of
equatorial disks.  It is then plausible that the scattering of light
from the central star by free electrons in the envelope or in the wind
might lead to a measurable polarization if the degree of rotational
flattening were sufficient and the orientation favorable.
Polarimetric observations are therefore a valuable asset in the study
of O-star winds, particularly as an indicator of the geometrical
distribution of the outflowing material.

The discovery that most O stars exhibit deep-seated spectroscopic
variability (Fullerton, Gies, \& Bolton 1996) opened up the
tantalizing possibility that photospheric activity may modulate the
stellar wind by acting at its base.  At nearly the same time,
exploratory observations of O stars with {\em IUE\/} revealed spectral
features (discrete absorption components, or DACs) accelerating
blueward across the profiles of UV resonance lines, as if tracing the
outflow of density perturbations in the wind (Prinja \& Howarth 1986).
Thus began a series of international multiwavelength observing
campaigns extending over several years (McDavid 1994), using {\em
IUE\/} together with simultaneous high resolution optical
spectroscopy, photometry, and polarimetry at various ground-based
observatories, in a search for ``the photospheric connection'':
observational evidence that wind variations are directly linked to
stellar surface activity.  As a sensitive technique for monitoring
conditions in the inner wind regions, polarimetry had a natural role
to play in these observing campaigns, and the purpose of this paper is
to presents the results. Some later followup observations are also
included.

\section{Collected Data}

The program stars and relevant data are listed in Table 1.  Tables
2--10 contain the results of a literature search for observations of
each star, followed by new observations which have not been previously
published.  The reference codes given in parentheses immediately after
the dates of observation are as follows: (1)~Mathewson et al.\ 1978;
(2a)~Coyne \& Gehrels 1966; (2b)~Serkowski, Gehrels, \& Wisniewski
1969; (2c)~Gehrels 1974; (3a)~Hayes 1975; (3b)~Hayes 1978; (3c)~Hayes
1984; (4)~Poeckert, Bastien, \& Landstreet 1979; (5)~Lupie
\& Nordsieck 1987; (6)~McDavid (McDonald Observatory, previously
unpublished); (7)~McDavid (Limber Observatory, previously
unpublished).

Where multiple measurements were available, the normalized Stokes
parameters $q$ and $u$, the degree $p$ of polarization, and the
equatorial position angle $\theta$ are the means of the given number
$n$ of individual measurements, and $dq$, $du$, $dp$, and $d\theta$
are the associated standard deviations.  Results are presented for $p$
and $\theta$ to help distinguish between variability in the degree of
polarization and variability of the position angle, even though these
parameters are not normally distributed (Clarke \& Stewart 1986).  For
single measurements (1, 2c) and some cases of only two measurements
(2a, 2b), $dq$, $du$, $dp$, and $d\theta$ are simply representative
error estimates, usually based on repeatability of frequent
observations of standard stars.  In many references $q$ and $u$ were
not given, so they were calculated as $q=p \cos 2\theta$, $u=p \sin
2\theta$, and $dq=du=dp$ for the sake of statistical completeness.  In
cases where $d\theta$ was missing, it was estimated as
$28\fdg65(dp/p)$.  The quantities $dpi$ and $d\theta i$ are estimates
of the uncertainty in a single measurement, which are sometimes
equivalent to the ``representative error estimates'' mentioned above
(1,2), sometimes derived from the residuals of the fits used in the
data reduction procedure (4,5), and sometimes calculated from photon
counting statistics (3, 6).  For the Limber Observatory data (7) a
single measurement and its error are the mean and standard deviation
of three repetitions.

\section{Long-Term Analysis: The Interstellar Component}

Simple inspection of Tables 2--10 is all that is necessary to see that
there is no compelling evidence for long-term variable polarization at
the 3$\sigma$ level for any of the program stars.  The complete data
set is far too heterogeneous and the sample sizes far too small to
expect any meaningful result from application of a rigorous analysis
of variance with F-test probabilities.  Relaxing the requirements, but
still in search of a more quantitative conclusion, Table 11 was
constructed as a statistical summary of the data on all of the program
stars.  This was done by calculating the mean and standard deviation
of $q$, $u$, $p$, and $\theta$ over the subsets of observations for each
of the stars (columns 2, 4, 6, and 8), averaging the standard
deviations from each subset (bracketed quantities in columns 3, 5, 7,
and 9), and averaging the error estimates for a single observation
(bracketed quantities in columns 10 and 11).

It is then quite clear that the overall scatter $dq$, $du$, $dp$, and
$d\theta$ from combining individual data subsets is never
substantially greater than would be expected on average as seen in
{\tt <$dq$\tt >}, {\tt <$du$\tt >}, {\tt <$dp$\tt >}, and {\tt
<$d\theta$\tt >} or, alternatively, as seen in {\tt <$dpi$\tt >} and
{\tt <$d\theta i$\tt >}.  The Grand Averages over all 5 passbands (in
rows beginning with ``GAV''), based on the expectation that any
variations should appear more or less in all filters, further
demonstrate the absence of any detectable variability.

Can we use Table 11 to make a quantitative statement about the
amplitude of genuine variability that would escape detection in this
study, as well as an amplitude of variability that would surely be
detected?  A conservative criterion for the break point would be the
average of the Grand Averages of $dp$ over all the stars in Table 11,
which is 0.07\%.  Smaller variations might be completely hidden, and
the threshold for 3$\sigma$ detection would be 0.21\%.  Compared to
state-of-the-art precision on the order of 0.02\% this may seem to be
a weak result, but considering that it was obtained from the
combination of at least seven different instrumental systems over a
span of 40 years, it is a perfectly realistic conclusion that none of
the program stars has varied substantially.  This absence of
variability constitutes strong evidence that the polarization is
predominantly interstellar.

To continue the test for an interstellar origin of the polarization, a
Serkowski law of the modified form $p(\lambda) = p_{max} \exp [-1.7
\lambda_{max} \ln^{2} (\lambda_{max} / \lambda)]$ found by Whittet et
al. (1992) was fitted to the $\ubvri$ data on each star by weighted
nonlinear least squares.  The latest Limber Observatory observations
were used because they comprise the most uniform data set.  The top
panels of Figures 1--9 show the results, including the fit parameters.
Since the best fit values of $\lambda_{max}$ generally fall within the
normal range of 0.45 $\micron$ to 0.8 $\micron$ (Serkowski, Methewson,
\& Ford 1975), this experiment can be taken to support the
interpretation of the polarization as interstellar.

The middle panels of Figures 1--9 show the polarization of the program
star and neighboring stars taken from the agglomeration catalog of
Heiles (1999), superimposed on the {\em IRAS\/} 100 $\micron$ survey
image, which traces the interstellar dust.  In most cases the position
angle of the program star is similar to those of its neighbors, as
would be expected if all the stars are polarized by selective
extinction due to a locally uniform alignment of the interstellar dust
grains (Aannestad \& Greenberg 1983).

The bottom panels of Figures 1--9 show the $\ubvri$ polarization of
the program star plotted as 1$\sigma$ confidence ellipses in the $q,u$
plane.  A solid line is drawn through the origin at the average of the
$\ubvri$ position angles.  For comparison, a dashed line is drawn
showing the best straight-line fit to the $\ubvri$ data points.  These
two lines will coincide if the position angle of the polarization is
independent of wavelength, which is expected if it is purely
interstellar.  The shaded region is centered on the mean polarization
position angle of the neighboring stars, extended by one standard
deviation on either side, and defines the region of agreement between
the position angle of the program star and those of the neighboring
stars.  The figures clearly show that the solid and dashed lines
generally match well, with the greatest differences occurring when the
degree of polarization is small and the position angle therefore
poorly defined.  In all cases except 19 Cep, which has only a
negligible mismatch, the solid and dashed lines fall within the shaded
regions.

To summarize, four arguments imply that the polarization of all 9
program stars is interstellar: (1) No long-term variability has ever
been detected. (2) The wavelength dependence of the polarization
follows a Serkowski law typical of interstellar polarization.  (3) The
position angle of the polarization is independent of wavelength. (4)
The position angle is consistent with those of neighboring stars.

In spite of this evidence against intrinsic polarization, it should be
mentioned that there have also been some investigations resulting in
its favor.  Hayes (1975) made rigorous statistical analyses on his
polarization measurements of several O stars with dense monitoring
coverage over periods of months and concluded that $\lambda$~Cep
(Hayes 1978) and $\alpha$~Cam (Hayes 1984) were variable at the
3$\sigma$ level.  He gave convincing proof of instrumental stability
to the photon counting limit of $\sim$ 0.02\%, which justified his
detection of variability at levels of 0.06\% to 0.08\%.  In a later
study Lupie \& Nordsieck (1987) found evidence for spectropolarimetric
variability of the same two stars with only a slightly lower level of
signal to noise.

As another example of the evidence for intrinsic polarization in
emission-line O stars, Harries and Howarth (1996) discovered a
spectropolarimetric change across the H$\alpha$ emission line of
$\zeta$~Pup, which implies a polarizing wind asymmetry.  Also, Ebbets
(1981) summarized the H$\alpha$ emission episodes of $\zeta$~Oph,
which seem to have escaped polarimetric observation but might well
have included polarization effects (Reid et al.~1993).  The x-ray
outburst of $\zeta$~Ori (Bergh{\"o}fer \& Schmitt 1994), which was
attributed to a propagating wind shock, also escaped polarimetric
observation.

\section{Short-Term Analysis: Hints of Cyclic Variability}

Most of the program stars were monitored at McDonald Observatory in
conjunction with simultaneous ultraviolet and optical spectroscopy
during several roughly week-long campaigns from 1986 through 1992.
These time series are identifiable in Tables 2--10 by entries with
large numbers of individual measurements, where only the averages are
given.  With a typical instrumental uncertainty of about 0.03\% for a
single observation, no polarization variability was detected at the
$3\sigma$ level for any of the series (McDavid 1998).  However, during
the campaign of October 1991, both 68~Cyg and $\xi$ Per showed
small-amplitude periodicities in polarization which are a priori
significant because they match those found by Kaper et al.~(1997) in
the simultaneous optical and ultraviolet spectra.  These variations
constitute evidence for intrinsic polarization of a form that could
have very easily escaped detection in the long-term monitoring
discussed earlier in this paper, but their reality should be regarded
as tentative.

A CLEANed Fourier periodogram (Roberts, Lehar, \& Dreher 1987) of the
polarization of 68~Cyg using 15 iterations at a gain of 0.9 shows
maximum power at 0.75 $\pm$ 0.03 ${\rm d^{-1}}$ (period 1.33 d), equal
to the frequency of both the H$\alpha$ and \ion{Si}{4} equivalent
widths.  The amplitude of this variation is about 0.045\% (see Figure
10).  Interestingly, the polarization is in antiphase with the
H$\alpha$ EW, exactly as expected in the corotating interaction region
(CIR) model applied to O-star winds (Cranmer \& Owocki 1996).  The
notes in Figure 11 give a qualitative plausibility argument for how
this might come about.  For a thorough theoretical treatment see
Harries (1999).  Assuming a radius of $14 {\rm R}_{\sun}$ and an
inclination of 90\arcdeg, the rotation period of 68~Cyg is 2.59 d.
This gives two cycles per rotation, also in agreement with the CIR
model with two diametrically opposite equatorial bright areas.

The case of $\xi$~Per is more complicated.  Kaper et al.~(1997) found
the H$\alpha$ and \ion{Si}{4} EWs to vary with a frequency of 0.50
$\pm$ 0.10 ${\rm d^{-1}}$ (period 2.0 d) and also noted the presence
of another H$\alpha$ frequency at 1.12 $\pm$ 0.03 ${\rm d^{-1}}$
(period 0.89 d), which they interpreted as a harmonic.  As shown in
Figure 12, the simultaneous polarimetry shows exactly this frequency,
with an amplitude of about 0.025\%.  Assuming a radius of $11 {\rm
R}_{\sun}$, the rotation period of $\xi$~Per is less than about 2.78
d, depending on the inclination.  It is possible, then, that the 2.0 d
period is the rotation period, in which case $\xi$~Per had only one
photospheric bright area and associated CIR structure at the time of
the observations.  This would result in one EW cycle per rotation, but
two polarization cycles, as observed.

\section{Conclusions}

This study has illustrated that intrinsic polarization in O stars is
clearly a rare phenomenon in spite of the similarity between
emission-line O stars and the cooler and less luminous Be stars, which
have well-known intrinsic polarization.  Both classes of stars have
significant mass loss in the form of variable winds, but the origin of
the variability is still poorly understood.  It seems a likely
conclusion that the long-lived flattened envelopes or disks
characteristic of Be stars almost never form around even the most
rapidly rotating O stars because the circumstellar regions of O stars
are more completely swept out by their stronger, luminosity-driven
winds.  The only indications of variable polarization found in the O
stars are weak suggestions of rapid cycles on the time scale of the
stellar rotation period (a few days at most) which appear to involve
wind density patterns that change aspect with rotation.

The O stars presented here deserve continued long-term polarization
monitoring for the type of outburst behavior seen in $\zeta$ Oph and
$\zeta$ Ori.  Further pursuit of the suspected rapid cyclic
variations, however, is likely to require dense time-series
polarimetry of unprecedented precision, with random errors not to
exceed 0.01\%.

\acknowledgments

I would like to take this opportunity to thank my friends and
coworkers on the ``O-Team'': Tom Bolton, Doug Gies, Alex Fullerton,
and Huib Henrichs, for their enthusiasm and encouragement.  We were
all inspired by the work of Andy Reid and Lex Kaper, who dispelled
some of our skepticism and showed us the value of our observations.  I
also appreciate the advice of Dietrich Baade and Ian Howarth, who both
took the time to read and comment on early versions of this paper.

I am grateful to the administration of McDonald Observatory for
generous allotments of observing time, as well as to Ed Dutchover and
the rest of the technical staff for maintaining the Breger polarimeter
in stable working order for year after year.

This research has made use of the SIMBAD database, operated at CDS,
Strasbourg, France.  I acknowledge the use of NASA's {\em SkyView\/}
facility (http://skyview.gsfc.nasa.gov) located at NASA Goddard Space
Flight Center.

\newpage

\begin{figure}
\caption{(top panel): Revised Serkowski law fit to the $\ubvri$
polarization of $\xi$~Per, showing values of the parameters $p_{max}$
and $\lambda_{max}$.  (middle panel): {\em IRAS\/} 100~$\micron$
image, centered on the program star, showing polarization of
neighboring stars.  North is up and East is left.  (bottom panel):
$q,u$ plot with analysis of the $\ubvri$ polarization position angles.
See \S 3 for further explanation.}
\end{figure}

\begin{figure}
\caption{Same as Fig.~1, only for $\alpha$ Cam.}
\end{figure}

\begin{figure}
\caption{Same as Fig.~1, only for $\lambda$ Ori.}
\end{figure}

\begin{figure}
\caption{Same as Fig.~1, only for $\zeta$ Ori.}
\end{figure}

\begin{figure}
\caption{Same as Fig.~1, only for $\zeta$ Oph.}
\end{figure}

\begin{figure}
\caption{Same as Fig.~1, only for 68 Cyg.}
\end{figure}

\begin{figure}
\caption{Same as Fig.~1, only for 19 Cep.}
\end{figure}

\begin{figure}
\caption{Same as Fig.~1, only for $\lambda$ Cep.}
\end{figure}

\begin{figure}
\caption{Same as Fig.~1, only for 10 Lac.}
\end{figure}

\begin{figure}
\caption{Correlations among cycles in the equivalent widths of \ion{Si}{4}
and H$\alpha$ and the polarization of 68~Cyg during the campaign of
October 1991, adapted from Kaper et al.\ (1997).}
\end{figure}

\begin{figure}
\caption{Illustration of the CIR model for a hot-star wind, with
annotations explaining qualitatively the expected relations among
\ion{Si}{4} and H$\alpha$ equivalent widths and the polarization
when viewed from different directions, adapted from Cranmer \& Owocki
(1996).}
\end{figure}

\begin{figure}
\caption{Correlations among cycles in the equivalent widths of \ion{Si}{4}
and H$\alpha$ and the polarization of $\xi$~Per during the campaign of
October 1991, adapted from Kaper et al.\ (1997).}
\end{figure}

\newpage

\begin{deluxetable}{rlclccccc}
\tablewidth{0pt}
\tablecolumns{9}
\tablecaption{Program Stars}
\tablehead{
\colhead{HD}&
\colhead{Name}&
\colhead{$V$\tablenotemark{a}}&
\colhead{Spectral}&
\colhead{$R$\tablenotemark{c}}&
\colhead{$T_{eff}$\tablenotemark{c}}&
\colhead{$\log L$\tablenotemark{c}}&
\colhead{Dist\tablenotemark{d}}&
\colhead{$v \sin i$\tablenotemark{e}}\\
&&(mag)&\phm{xxxx}Type\tablenotemark{b}&($R_{\sun}$)&($K$)&($L_{\sun}$)&(pc)&(${\rm km~s^{-1}}$)
}

\startdata

24912&$\xi$ Per&4.04&O7.5 III(n)((f))&11&36\,000&5.3&\phn457&213\\

30614&$\alpha$ Cam&4.29&O9.5 Ia&22& 29\,900&5.5&1088& 129\\

36861&$\lambda$ Ori&3.66&O8 III((f))&12&35\,000&5.3&\phn372&\phn74\\

37742&$\zeta$ Ori&1.75&O9.7 Ib&29&30\,000&5.8&\phn332&124\\

149757&$\zeta$ Oph&2.56&O9.5 V&\phn8&34\,000&4.9&\phn163&372\\ 

203064&68 Cyg&5.00&O7.5 III:n((f))&14&36\,000&5.5&\phn660&305\\

209975&19 Cep&5.11&O9.5 Ib&18&30\,200&5.4&1090&\phn95\\

210839&$\lambda$ Cep&5.04&O6 I(n)fp&17&42\,000&5.9&\phn550&219\\

214680&10 Lac&4.88&O9 V&\phn9&38\,000&5.1&\phn704&\phn35\\

\enddata

\tablerefs{
(a) Hoffleit \& Jaschek 1982;
(b) Walborn 1972, 1973;
(c) Howarth \& Prinja 1989;
(d) Heiles 1999;
(e) Howarth et al.\ 1997
}

\end{deluxetable}

\begin{deluxetable}{lcccccc}
\tablecolumns{7}
\tablewidth{0pt}
\tablecaption{$\xi$ Per}
\tablehead{
\colhead{Year(Source)} &
\colhead{$q\/$/$dq$} &
\colhead{$u\/$/$du$} &
\colhead{$p\/$/$dp$} &
\colhead{$\theta\/$/$d\theta$} &
\colhead{$dpi$} &
\colhead{$d \theta i$} \\
\colhead{Filter($n$)} &
\colhead{(\%)} &
\colhead{(\%)} &
\colhead{(\%)} &
\colhead{($\arcdeg$)} &
\colhead{(\%)} &
\colhead{($\arcdeg$)}
}

\startdata

\sidehead{1949--1958(1)}

$B$(1) & -1.12/0.20 & -0.65/0.20 & \phm{-}1.29/0.20 & 105.0/\phn4.4 & 0.20 & \phn4.4\\

\sidehead{1966(2b)}

$U$(2) & -0.96/0.10 & -0.66/0.10 & \phm{-}1.16/0.10 & 107.2/\phn2.0 & 0.10 & \phn2.0\\

$B$(2) & -1.09/0.10 & -0.82/0.10 & \phm{-}1.37/0.10 & 108.5/\phn2.0 & 0.10 & \phn2.0\\

$V$(2) & -1.15/0.10 & -0.89/0.10 & \phm{-}1.46/0.10 & 108.9/\phn2.0 & 0.10 & \phn2.0\\

$R$(2) & -1.19/0.10 & -0.87/0.10 & \phm{-}1.47/0.10 & 108.1/\phn2.0 & 0.10 & \phn2.0\\

$I$(2) & -1.00/0.10 & -0.75/0.10 & \phm{-}1.25/0.10 & 108.5/\phn2.0 & 0.10 & \phn2.0\\

\sidehead{1989.79(6)}

$U$(6) & -0.86/0.05 & -0.60/0.06 & \phm{-}1.05/0.03 & 107.4/\phn1.9 & 0.04 & \phn1.1\\

$B$(22)& -0.98/0.03 & -0.74/0.02 & \phm{-}1.23/0.03 & 108.4/\phn0.5 & 0.03 & \phn0.9\\

$V$(6) & -1.11/0.02 & -0.83/0.06 & \phm{-}1.39/0.03 & 108.4/\phn1.1 & 0.03 & \phn1.3\\

$R$(6) & -1.12/0.03 & -0.82/0.02 & \phm{-}1.39/0.02 & 108.2/\phn0.5 & 0.02 & \phn0.5\\

$I$(6) & -0.99/0.06 & -0.74/0.08 & \phm{-}1.24/0.05 & 108.4/\phn2.0 & 0.05 & \phn1.1\\

\sidehead{1991.82(6)}

$V$(15)& -1.13/0.03 & -0.85/0.02 & \phm{-}1.41/0.02 & 108.5/\phn0.6 & 0.02 & \phn0.5\\

\sidehead{1996.13(7)}

$U$(1) & -0.79/0.14 & -0.59/0.14 & \phm{-}0.99/0.20 & 108.2/\phn1.0 & 0.20 & \phn1.0\\

$B$(1) & -0.96/0.01 & -0.79/0.11 & \phm{-}1.24/0.06 & 109.7/\phn2.1 & 0.06 & \phn2.1\\

$V$(1) & -0.99/0.12 & -0.79/0.06 & \phm{-}1.28/0.06 & 109.4/\phn2.7 & 0.06 & \phn2.7\\

$R$(1) & -1.14/0.12 & -0.90/0.03 & \phm{-}1.46/0.09 & 109.2/\phn1.9 & 0.09 & \phn1.9\\

$I$(1) & -1.07/0.09 & -0.76/0.14 & \phm{-}1.32/0.01 & 107.8/\phn3.6 & 0.01 & \phn3.6\\

\sidehead{1997.04(7)}

$U$(4) & -0.90/0.02 & -0.56/0.08 & \phm{-}1.06/0.04 & 106.1/\phn1.8 & 0.10 & \phn1.9\\

$B$(4) & -0.95/0.05 & -0.65/0.06 & \phm{-}1.16/0.07 & 107.2/\phn0.8 & 0.06 & \phn1.7\\

$V$(4) & -1.07/0.05 & -0.74/0.10 & \phm{-}1.30/0.09 & 107.2/\phn1.4 & 0.05 & \phn1.1\\

$R$(4) & -1.06/0.05 & -0.68/0.06 & \phm{-}1.27/0.07 & 106.4/\phn0.6 & 0.11 & \phn3.4\\

$I$(4) & -1.06/0.07 & -0.65/0.07 & \phm{-}1.24/0.05 & 105.7/\phn2.1 & 0.11 & \phn2.0\\

\enddata
\end{deluxetable}

\begin{deluxetable}{lcccccc}
\tablecolumns{7}
\tablewidth{0pt}
\tablecaption{$\alpha$ Cam}
\tablehead{
\colhead{Year(Source)} &
\colhead{$q\/$/$dq$} &
\colhead{$u\/$/$du$} &
\colhead{$p\/$/$dp$} &
\colhead{$\theta\/$/$d\theta$} &
\colhead{$dpi$} &
\colhead{$d \theta i$} \\
\colhead{Filter($n$)} &
\colhead{(\%)} &
\colhead{(\%)} &
\colhead{(\%)} &
\colhead{($\arcdeg$)} &
\colhead{(\%)} &
\colhead{($\arcdeg$)}
}

\startdata

\sidehead{1949--1958(1)}

$B$(1) & \phm{-}0.35/0.20 & -1.62/0.20 & \phm{-}1.66/0.20 & 141.0/\phn3.4 & 0.20 & \phn3.4\\

\sidehead{1963--1965(2a)}

$U$(2) & \phm{-}0.20/0.08 & -1.67/0.08 & \phm{-}1.68/0.08 & 138.4/\phn2.7 & 0.08 & \phn0.9\\

$B$(2) & \phm{-}0.25/0.08 & -1.74/0.08 & \phm{-}1.76/0.08 & 139.1/\phn0.9 & 0.08 & \phn0.9\\

$V$(2) & \phm{-}0.22/0.08 & -1.82/0.08 & \phm{-}1.83/0.08 & 138.4/\phn0.9 & 0.08 & \phn0.9\\

$R$(2) & \phm{-}0.10/0.08 & -1.44/0.08 & \phm{-}1.44/0.08 & 136.9/\phn0.9 & 0.08 & \phn0.9\\

$I$(2) & \phm{-}0.04/0.08 & -1.13/0.08 & \phm{-}1.13/0.08 & 136.0/\phn0.9 & 0.08 & \phn0.9\\

\sidehead{1978.8(3c)}

$B$(41) & \phm{-}0.16/0.08 & -1.54/0.08 & \phm{-}1.55/0.08 & 138.0/\phn1.0 & 0.02 & \phn0.4\\

\sidehead{1979--1982(5)}

$V$(8) & \phm{-}0.01/0.07 & -1.46/0.07 & \phm{-}1.46/0.07 & 135.2/\phn1.8 & 0.02 & \phn0.4\\

$R$(8) & \phm{-}0.02/0.15 & -1.59/0.15 & \phm{-}1.59/0.13 & 135.3/\phn1.7 & 0.04 & \phn1.0\\

$I$(8) & \phm{-}0.04/0.05 & -1.41/0.05 & \phm{-}1.41/0.05 & 135.9/\phn2.6 & 0.04 & \phn1.2\\

\sidehead{1997.10(7)}

$U$(4) & \phm{-}0.16/0.12 & -1.37/0.16 & \phm{-}1.39/0.15 & 138.4/\phn2.7 & 0.14 & \phn2.7\\

$B$(4) & \phm{-}0.15/0.08 & -1.51/0.10 & \phm{-}1.52/0.09 & 137.9/\phn1.6 & 0.06 & \phn1.4\\

$V$(4) & \phm{-}0.15/0.08 & -1.47/0.09 & \phm{-}1.48/0.09 & 137.9/\phn1.7 & 0.07 & \phn1.3\\

$R$(4) & \phm{-}0.14/0.10 & -1.39/0.16 & \phm{-}1.40/0.16 & 137.9/\phn2.0 & 0.06 & \phn1.7\\

$I$(4) & \phm{-}0.09/0.12 & -1.16/0.12 & \phm{-}1.18/0.12 & 137.4/\phn2.7 & 0.06 & \phn4.0\\

\enddata
\end{deluxetable}

\begin{deluxetable}{lcccccc}
\tablecolumns{7}
\tablewidth{0pt}
\tablecaption{$\lambda$ Ori}
\tablehead{
\colhead{Year(Source)} &
\colhead{$q\/$/$dq$} &
\colhead{$u\/$/$du$} &
\colhead{$p\/$/$dp$} &
\colhead{$\theta\/$/$d\theta$} &
\colhead{$dpi$} &
\colhead{$d \theta i$} \\
\colhead{Filter($n$)} &
\colhead{(\%)} &
\colhead{(\%)} &
\colhead{(\%)} &
\colhead{($\arcdeg$)} &
\colhead{(\%)} &
\colhead{($\arcdeg$)}
}

\startdata

\sidehead{1949--1958(1)}

$B$(1) & -0.16/0.20 & -0.20/0.20 & \phm{-}0.26/0.20 & 116.0/22.0 & 0.20 & 22.0\\

\sidehead{1974.54(3a)}

$B$(12)& -0.23/0.02 & -0.11/0.02 & \phm{-}0.26/0.02 & 102.7/\phn2.2 & 0.02 & \phn2.2\\

\sidehead{1992.86(6)}

$U$(3) & -0.22/0.02 & -0.10/0.06 & \phm{-}0.25/0.04 & 101.8/\phn5.0 & 0.03 & \phn3.2\\

$B$(3) & -0.25/0.03 & -0.14/0.03 & \phm{-}0.29/0.02 & 105.0/\phn3.0 & 0.02 & \phn1.8\\

$V$(16)& -0.27/0.04 & -0.12/0.04 & \phm{-}0.30/0.04 & 102.2/\phn3.7 & 0.03 & \phn2.8\\

$R$(3) & -0.23/0.07 & -0.10/0.02 & \phm{-}0.25/0.07 & 102.2/\phn1.9 & 0.02 & \phn2.3\\

$I$(3) & -0.23/0.02 & -0.05/0.04 & \phm{-}0.24/0.02 & \phn96.6/\phn5.6 & 0.02 & \phn2.2\\

\sidehead{1997.10(7)}

$U$(4) & -0.25/0.03 & -0.17/0.09 & \phm{-}0.31/0.07 & 106.2/\phn7.0 & 0.04 & \phn7.6\\

$B$(4) & -0.29/0.02 & -0.22/0.08 & \phm{-}0.38/0.06 & 108.9/\phn5.5 & 0.04 & \phn9.1\\

$V$(4) & -0.24/0.04 & -0.16/0.06 & \phm{-}0.31/0.04 & 106.7/\phn5.7 & 0.03 & 10.7\\

$R$(4) & -0.27/0.03 & -0.13/0.09 & \phm{-}0.32/0.05 & 102.0/\phn8.0 & 0.04 & \phn8.4\\

$I$(4) & -0.27/0.07 & -0.07/0.12 & \phm{-}0.32/0.07 & \phn97.5/10.2 & 0.07 & 12.0\\

\enddata
\end{deluxetable}

\begin{deluxetable}{lcccccc}
\tablecolumns{7}
\tablewidth{0pt}
\tablecaption{$\zeta$ Ori}
\tablehead{
\colhead{Year(Source)} &
\colhead{$q\/$/$dq$} &
\colhead{$u\/$/$du$} &
\colhead{$p\/$/$dp$} &
\colhead{$\theta\/$/$d\theta$} &
\colhead{$dpi$} &
\colhead{$d \theta i$} \\
\colhead{Filter($n$)} &
\colhead{(\%)} &
\colhead{(\%)} &
\colhead{(\%)} &
\colhead{($\arcdeg$)} &
\colhead{(\%)} &
\colhead{($\arcdeg$)}
}

\startdata

\sidehead{1949--1958(1)}

$B$(1) & -0.23/0.12 & \phm{-}0.06/0.12 & \phm{-}0.24/0.12 & \phn83.0/14.3 & 0.12 & 14.3\\

\sidehead{1979--1982(5)}

$V$(3) & -0.38/0.02 & \phm{-}0.11/0.02 & \phm{-}0.40/0.02 & \phn81.9/\phn0.5 & 0.02 & \phn1.4\\

\sidehead{1992.86(6)}

$U$(3) & -0.15/0.02 & \phm{-}0.10/0.01 & \phm{-}0.18/0.02 & \phn72.8/\phn2.3 & 0.01 & \phn2.1\\

$B$(3) & -0.12/0.04 & \phm{-}0.14/0.02 & \phm{-}0.19/0.01 & \phn65.7/\phn6.7 & 0.01 & \phn1.4\\

$V$(15)& -0.19/0.04 & \phm{-}0.12/0.04 & \phm{-}0.23/0.03 & \phn73.5/\phn5.5 & 0.03 & \phn3.6\\

$R$(3) & -0.21/0.04 & \phm{-}0.08/0.01 & \phm{-}0.22/0.04 & \phn79.6/\phn1.6 & 0.01 & \phn1.2\\

$I$(3) & -0.11/0.03 & \phm{-}0.19/0.03 & \phm{-}0.22/0.04 & \phn59.8/\phn3.1 & 0.02 & \phn2.9\\

\sidehead{1997.10(7)}

$U$(4) & -0.23/0.10 & \phm{-}0.03/0.11 & \phm{-}0.28/0.08 & \phn81.8/13.6 & 0.10 & 11.2\\

$B$(4) & -0.21/0.17 & \phm{-}0.00/0.05 & \phm{-}0.27/0.12 & \phn89.0/16.8 & 0.09 & 25.1\\

$V$(4) & -0.19/0.14 & \phm{-}0.07/0.09 & \phm{-}0.25/0.12 & \phn84.0/15.7 & 0.07 & 25.7\\

$R$(4) & -0.20/0.09 & \phm{-}0.09/0.12 & \phm{-}0.25/0.11 & \phn81.2/12.9 & 0.10 & 12.6\\

$I$(4) & -0.23/0.15 & \phm{-}0.03/0.08 & \phm{-}0.28/0.13 & \phn86.9/13.3 & 0.05 & 17.5\\

\enddata
\end{deluxetable}

\begin{deluxetable}{lcccccc}
\tablecolumns{7}
\tablewidth{0pt}
\tablecaption{$\zeta$ Oph}
\tablehead{
\colhead{Year(Source)} &
\colhead{$q\/$/$dq$} &
\colhead{$u\/$/$du$} &
\colhead{$p\/$/$dp$} &
\colhead{$\theta\/$/$d\theta$} &
\colhead{$dpi$} &
\colhead{$d \theta i$} \\
\colhead{Filter($n$)} &
\colhead{(\%)} &
\colhead{(\%)} &
\colhead{(\%)} &
\colhead{($\arcdeg$)} &
\colhead{(\%)} &
\colhead{($\arcdeg$)}
}

\startdata

\sidehead{1949--1958(1)}

$B$(1) & -0.34/0.12 & -1.37/0.12 & \phm{-}1.41/0.12 & 128.0/\phn2.4 & 0.12 & \phn2.4\\

\sidehead{1964--1966(2c)}

$U$(1) & -0.28/0.01 & -1.08/0.01 & \phm{-}1.12/0.01 & 127.8/\phn0.3 & 0.01 & \phn0.3\\

$B$(1) & -0.32/0.02 & -1.28/0.02 & \phm{-}1.32/0.02 & 127.9/\phn0.5 & 0.02 & \phn0.5\\

$V$(1) & -0.39/0.01 & -1.43/0.01 & \phm{-}1.48/0.01 & 127.4/\phn0.1 & 0.01 & \phn0.1\\

$R$(1) & -0.40/0.01 & -1.44/0.01 & \phm{-}1.49/0.01 & 127.2/\phn0.4 & 0.01 & \phn0.4\\

$I$(1) & -0.34/0.01 & -1.24/0.01 & \phm{-}1.29/0.01 & 127.3/\phn0.2 & 0.01 & \phn0.2\\

\sidehead{1974.54(3a)}

$B$(12)& -0.33/0.02 & -1.25/0.02 & \phm{-}1.29/0.02 & 127.5/\phn0.4 & 0.02 & \phn0.4\\

\sidehead{1976.4(4)}

$U$(2) & -0.24/0.12 & -0.97/0.12 & \phm{-}1.00/0.12 & 128.2/\phn0.7 & 0.02 & \phn0.6\\

$B$(2) & -0.37/0.04 & -1.13/0.04 & \phm{-}1.17/0.04 & 127.4/\phn0.2 & 0.02 & \phn0.5\\

$I$(1) & -0.38/0.01 & -1.17/0.01 & \phm{-}1.23/0.01 & 126.1/\phn0.2 & 0.01 & \phn0.2\\

\sidehead{1981--1982(5)}

$V$(6) & -0.56/0.04 & -1.31/0.04 & \phm{-}1.43/0.04 & 123.4/\phn1.1 & 0.03 & \phn0.6\\

$R$(6) & -0.60/0.04 & -1.34/0.04 & \phm{-}1.47/0.04 & 123.0/\phn0.5 & 0.04 & \phn1.0\\

$I$(6) & -0.55/0.04 & -1.29/0.05 & \phm{-}1.40/0.04 & 123.4/\phn1.4 & 0.04 & \phn1.1\\

\sidehead{1989.32(6)}

$B$(5) & -0.41/0.03 & -1.28/0.05 & \phm{-}1.35/0.06 & 126.2/\phn0.2 & 0.03 & \phn0.6\\

$V$(17)& -0.50/0.03 & -1.41/0.03 & \phm{-}1.50/0.04 & 125.3/\phn0.6 & 0.04 & \phn0.6\\

\sidehead{1992.71(6)}

$U$(1) & -0.26/0.04 & -1.01/0.04 & \phm{-}1.05/0.04 & 127.8/\phn1.0 & 0.04 & \phn1.0\\

$B$(1) & -0.28/0.02 & -1.34/0.02 & \phm{-}1.37/0.02 & 129.0/\phn0.4 & 0.02 & \phn0.4\\

$V$(1) & -0.56/0.02 & -1.37/0.02 & \phm{-}1.48/0.02 & 124.0/\phn0.4 & 0.02 & \phn0.4\\

$R$(1) & -0.50/0.02 & -1.40/0.02 & \phm{-}1.49/0.02 & 125.1/\phn0.3 & 0.02 & \phn0.3\\

$I$(1) & -0.56/0.04 & -1.52/0.04 & \phm{-}1.62/0.04 & 124.8/\phn0.7 & 0.04 & \phn0.7\\

\sidehead{1996.35(7)}

$U$(3) & -0.28/0.02 & -1.16/0.05 & \phm{-}1.19/0.06 & 128.3/\phn0.1 & 0.10 & \phn1.7\\

$B$(3) & -0.36/0.05 & -1.26/0.03 & \phm{-}1.31/0.02 & 127.1/\phn1.2 & 0.04 & \phn1.1\\

$V$(3) & -0.41/0.02 & -1.44/0.01 & \phm{-}1.50/0.01 & 127.1/\phn0.3 & 0.05 & \phn1.4\\

$R$(3) & -0.45/0.08 & -1.32/0.05 & \phm{-}1.40/0.07 & 125.7/\phn1.4 & 0.06 & \phn1.0\\

$I$(3) & -0.41/0.07 & -1.29/0.06 & \phm{-}1.36/0.04 & 126.0/\phn1.7 & 0.08 & \phn2.6\\

\sidehead{1997.51(7)}

$U$(3) & -0.16/0.04 & -1.04/0.07 & \phm{-}1.05/0.07 & 130.7/\phn1.3 & 0.06 & \phn2.0\\

$B$(3) & -0.31/0.05 & -1.25/0.02 & \phm{-}1.29/0.03 & 128.0/\phn0.9 & 0.05 & \phn1.0\\

$V$(3) & -0.33/0.07 & -1.39/0.04 & \phm{-}1.44/0.02 & 128.2/\phn1.5 & 0.06 & \phn1.2\\

$R$(3) & -0.31/0.04 & -1.43/0.08 & \phm{-}1.46/0.09 & 128.9/\phn0.4 & 0.05 & \phn0.8\\

$I$(3) & -0.43/0.03 & -1.24/0.09 & \phm{-}1.31/0.09 & 125.5/\phn0.8 & 0.08 & \phn1.5\\

\enddata
\end{deluxetable}

\begin{deluxetable}{lcccccc}
\tablecolumns{7}
\tablewidth{0pt}
\tablecaption{68 Cyg}
\tablehead{
\colhead{Year(Source)} &
\colhead{$q\/$/$dq$} &
\colhead{$u\/$/$du$} &
\colhead{$p\/$/$dp$} &
\colhead{$\theta\/$/$d\theta$} &
\colhead{$dpi$} &
\colhead{$d \theta i$} \\
\colhead{Filter($n$)} &
\colhead{(\%)} &
\colhead{(\%)} &
\colhead{(\%)} &
\colhead{($\arcdeg$)} &
\colhead{(\%)} &
\colhead{($\arcdeg$)}
}

\startdata

\sidehead{1949--1958(1)}

$B$(1) & -0.13/0.20 & \phm{-}0.35/0.20 & \phm{-}0.37/0.20 & \phn55.0/15.5 & 0.20 & 15.5\\

\sidehead{1986.65(6)}

$B$(3) & -0.28/0.02 & \phm{-}0.49/0.02 & \phm{-}0.57/0.02 & \phn60.0/\phn0.8 & 0.01 & \phn0.4\\

$V$(10) & -0.27/0.02 & \phm{-}0.54/0.02 & \phm{-}0.61/0.02 & \phn58.4/\phn1.3 & 0.01 & \phn0.6\\

\sidehead{1987.68(6)}

$V$(20) & -0.30/0.03 & \phm{-}0.50/0.02 & \phm{-}0.59/0.02 & \phn60.3/\phn1.5 & 0.03 & \phn1.2\\

\sidehead{1991.82(6)}

$U$(3) & -0.32/0.02 & \phm{-}0.40/0.04 & \phm{-}0.51/0.04 & \phn64.1/\phn1.0 & 0.01 & \phn0.6\\

$B$(3) & -0.38/0.05 & \phm{-}0.44/0.08 & \phm{-}0.59/0.04 & \phn65.3/\phn4.5 & 0.01 & \phn0.5\\

$V$(17) & -0.40/0.02 & \phm{-}0.50/0.03 & \phm{-}0.64/0.02 & \phn64.3/\phn1.4 & 0.03 & \phn1.7\\

$R$(3) & -0.40/0.04 & \phm{-}0.44/0.04 & \phm{-}0.60/0.05 & \phn66.1/\phn1.3 & 0.01 & \phn0.6\\

$I$(3) & -0.42/0.07 & \phm{-}0.40/0.02 & \phm{-}0.58/0.06 & \phn68.0/\phn2.2 & 0.04 & \phn2.0\\

\sidehead{1996.54(7)}

$U$(4) & -0.28/0.02 & \phm{-}0.52/0.05 & \phm{-}0.61/0.03 & \phn59.3/\phn1.8 & 0.14 & \phn7.5\\

$B$(4) & -0.34/0.03 & \phm{-}0.48/0.03 & \phm{-}0.60/0.04 & \phn62.6/\phn1.0 & 0.08 & \phn2.8\\

$V$(4) & -0.36/0.04 & \phm{-}0.52/0.06 & \phm{-}0.64/0.05 & \phn62.4/\phn2.4 & 0.10 & \phn1.2\\

$R$(4) & -0.37/0.05 & \phm{-}0.54/0.01 & \phm{-}0.66/0.03 & \phn62.0/\phn2.2 & 0.07 & \phn3.1\\

$I$(4) & -0.31/0.06 & \phm{-}0.46/0.11 & \phm{-}0.57/0.12 & \phn61.9/\phn1.6 & 0.11 & \phn9.1\\

\sidehead{1997.53(7)}

$U$(3) & -0.23/0.09 & \phm{-}0.50/0.08 & \phm{-}0.56/0.07 & \phn57.5/\phn4.9 & 0.13 & \phn5.6\\

$B$(3) & -0.32/0.02 & \phm{-}0.45/0.03 & \phm{-}0.56/0.02 & \phn62.7/\phn1.8 & 0.07 & \phn3.1\\

$V$(3) & -0.39/0.05 & \phm{-}0.51/0.06 & \phm{-}0.65/0.03 & \phn64.1/\phn3.2 & 0.06 & \phn3.5\\

$R$(3) & -0.34/0.04 & \phm{-}0.47/0.07 & \phm{-}0.58/0.05 & \phn62.9/\phn2.8 & 0.05 & \phn2.2\\

$I$(3) & -0.34/0.06 & \phm{-}0.49/0.08 & \phm{-}0.60/0.07 & \phn62.0/\phn2.9 & 0.13 & \phn3.6\\

\sidehead{1997.66(7)}

$V$(32) & -0.31/0.06 & \phm{-}0.50/0.06 & \phm{-}0.59/0.06 & \phn60.9/\phn2.3 & 0.05 & \phn2.8\\

\enddata
\end{deluxetable}

\begin{deluxetable}{lcccccc}
\tablecolumns{7}
\tablewidth{0pt}
\tablecaption{19 Cep}
\tablehead{
\colhead{Year(Source)} &
\colhead{$q\/$/$dq$} &
\colhead{$u\/$/$du$} &
\colhead{$p\/$/$dp$} &
\colhead{$\theta\/$/$d\theta$} &
\colhead{$dpi$} &
\colhead{$d \theta i$} \\
\colhead{Filter($n$)} &
\colhead{(\%)} &
\colhead{(\%)} &
\colhead{(\%)} &
\colhead{($\arcdeg$)} &
\colhead{(\%)} &
\colhead{($\arcdeg$)}
}

\startdata

\sidehead{1949--1958(1)}

$B$(1) & -0.79/0.20 & \phm{-}0.66/0.20 & \phm{-}1.03/0.20 & \phn70.0/\phn5.6 & 0.20 & \phn5.6\\

\sidehead{1986.65(6)}

$B$(2) & -0.96/0.02 & \phm{-}0.78/0.03 & \phm{-}1.23/0.00 & \phn70.5/\phn0.9 & 0.01 & \phn0.2\\

$V$(6) & -0.88/0.04 & \phm{-}0.78/0.07 & \phm{-}1.18/0.02 & \phn69.2/\phn1.9 & 0.01 & \phn0.3\\

\sidehead{1996.59(7)}

$U$(4) & -0.88/0.12 & \phm{-}0.67/0.10 & \phm{-}1.12/0.15 & \phn71.4/\phn0.2 & 0.12 & \phn2.5\\

$B$(4) & -1.01/0.09 & \phm{-}0.60/0.06 & \phm{-}1.18/0.10 & \phn74.7/\phn1.1 & 0.06 & \phn2.0\\

$V$(4) & -0.90/0.08 & \phm{-}0.57/0.08 & \phm{-}1.07/0.09 & \phn73.9/\phn2.1 & 0.10 & \phn1.4\\

$R$(4) & -0.83/0.04 & \phm{-}0.55/0.06 & \phm{-}1.01/0.04 & \phn73.4/\phn1.7 & 0.04 & \phn4.2\\

$I$(4) & -0.73/0.10 & \phm{-}0.45/0.04 & \phm{-}0.86/0.10 & \phn74.2/\phn1.9 & 0.15 & \phn3.8\\

\enddata
\end{deluxetable}

\begin{deluxetable}{lcccccc}
\tablecolumns{7}
\tablewidth{0pt}
\tablecaption{$\lambda$ Cep}
\tablehead{
\colhead{Year(Source)} &
\colhead{$q\/$/$dq$} &
\colhead{$u\/$/$du$} &
\colhead{$p\/$/$dp$} &
\colhead{$\theta\/$/$d\theta$} &
\colhead{$dpi$} &
\colhead{$d \theta i$} \\
\colhead{Filter($n$)} &
\colhead{(\%)} &
\colhead{(\%)} &
\colhead{(\%)} &
\colhead{($\arcdeg$)} &
\colhead{(\%)} &
\colhead{($\arcdeg$)}
}

\startdata

\sidehead{1949--1958(1)}

$B$(1) & -0.55/0.20 & \phm{-}1.03/0.20 & \phm{-}1.17/0.20 & \phn59.0/\phn4.9 & 0.20 & \phn4.9\\

\sidehead{1974--1976(3b)}

$B$(74)& -0.48/0.06 & \phm{-}1.13/0.06 & \phm{-}1.23/0.06 & \phn56.4/\phn0.9 & 0.02 & \phn0.5\\

\sidehead{1979--1982(5)}

$V$(8) & -0.52/0.04 & \phm{-}1.10/0.04 & \phm{-}1.22/0.04 & \phn57.7/\phn1.4 & 0.03 & \phn0.6\\

$R$(8) & -0.63/0.05 & \phm{-}1.17/0.05 & \phm{-}1.33/0.05 & \phn59.1/\phn1.4 & 0.06 & \phn1.5\\

$I$(8) & -0.48/0.07 & \phm{-}1.06/0.07 & \phm{-}1.16/0.07 & \phn57.2/\phn2.3 & 0.05 & \phn1.6\\

\sidehead{1986.65(6)}

$B$(2) & -0.47/0.01 & \phm{-}1.11/0.01 & \phm{-}1.20/0.01 & \phn56.5/\phn0.2 & 0.01 & \phn0.2\\

$V$(5) & -0.41/0.02 & \phm{-}1.14/0.05 & \phm{-}1.21/0.05 & \phn55.0/\phn0.7 & 0.02 & \phn0.4\\

\sidehead{1987.68(6)}

$V$(7) & -0.48/0.01 & \phm{-}1.09/0.03 & \phm{-}1.19/0.02 & \phn57.0/\phn0.4 & 0.02 & \phn0.6\\

\sidehead{1989.79(6)}

$U$(6) & -0.58/0.09 & \phm{-}1.02/0.09 & \phm{-}1.17/0.06 & \phn59.9/\phn2.6 & 0.08 & \phn1.9\\

$B$(24)& -0.49/0.05 & \phm{-}1.10/0.03 & \phm{-}1.21/0.04 & \phn56.9/\phn1.0 & 0.03 & \phn0.7\\

$V$(6) & -0.51/0.03 & \phm{-}1.01/0.19 & \phm{-}1.14/0.18 & \phn58.7/\phn2.0 & 0.05 & \phn1.2\\

$R$(6) & -0.44/0.04 & \phm{-}1.07/0.03 & \phm{-}1.16/0.02 & \phn56.2/\phn1.1 & 0.03 & \phn0.8\\

$I$(6) & -0.33/0.12 & \phm{-}0.88/0.10 & \phm{-}0.95/0.10 & \phn55.2/\phn3.7 & 0.07 & \phn2.2\\

\sidehead{1991.82(6)}

$V$(16)& -0.46/0.07 & \phm{-}1.09/0.04 & \phm{-}1.18/0.04 & \phn56.3/\phn1.7 & 0.04 & \phn1.1\\

\sidehead{1996.54(7)}

$U$(4) & -0.44/0.10 & \phm{-}0.99/0.10 & \phm{-}1.09/0.12 & \phn56.9/\phn2.1 & 0.12 & \phn2.7\\

$B$(4) & -0.54/0.08 & \phm{-}1.07/0.07 & \phm{-}1.20/0.08 & \phn58.4/\phn1.9 & 0.06 & \phn1.5\\

$V$(4) & -0.53/0.09 & \phm{-}1.09/0.05 & \phm{-}1.21/0.08 & \phn58.1/\phn1.5 & 0.06 & \phn2.3\\

$R$(4) & -0.51/0.08 & \phm{-}1.05/0.03 & \phm{-}1.17/0.03 & \phn57.9/\phn2.0 & 0.07 & \phn1.5\\

$I$(4) & -0.42/0.09 & \phm{-}0.90/0.09 & \phm{-}1.00/0.11 & \phn57.2/\phn2.0 & 0.14 & \phn3.6\\

\enddata
\end{deluxetable}

\begin{deluxetable}{lcccccc}
\tablecolumns{7}
\tablewidth{0pt}
\tablecaption{10 Lac}
\tablehead{
\colhead{Year(Source)} &
\colhead{$q\/$/$dq$} &
\colhead{$u\/$/$du$} &
\colhead{$p\/$/$dp$} &
\colhead{$\theta\/$/$d\theta$} &
\colhead{$dpi$} &
\colhead{$d \theta i$} \\
\colhead{Filter($n$)} &
\colhead{(\%)} &
\colhead{(\%)} &
\colhead{(\%)} &
\colhead{($\arcdeg$)} &
\colhead{(\%)} &
\colhead{($\arcdeg$)}
}

\startdata

\sidehead{1949--1958(1)}

$B$(1) & -0.49/0.12 & \phm{-}0.00/0.12 & \phm{-}0.49/0.12 & \phn90.0/\phn7.0 & 0.12 & \phn7.0\\

\sidehead{1992.86(6)}

$U$(3) & -0.45/0.08 & \phm{-}0.08/0.03 & \phm{-}0.46/0.08 & \phn84.8/\phn2.0 & 0.05 & \phn3.2\\

$B$(3) & -0.50/0.04 & \phm{-}0.10/0.06 & \phm{-}0.52/0.02 & \phn84.1/\phn3.7 & 0.04 & \phn1.9\\

$V$(18)& -0.52/0.03 & \phm{-}0.04/0.03 & \phm{-}0.52/0.03 & \phn87.9/\phn1.6 & 0.04 & \phn1.2\\ 

$R$(3) & -0.52/0.06 & \phm{-}0.01/0.02 & \phm{-}0.52/0.06 & \phn89.6/\phn1.3 & 0.04 & \phn2.0\\

$I$(3) & -0.47/0.05 & \phm{-}0.03/0.04 & \phm{-}0.47/0.05 & \phn88.3/\phn2.1 & 0.04 & \phn2.4\\

\sidehead{1996.59(7)}

$U$(4) & -0.39/0.03 & -0.01/0.10 & \phm{-}0.41/0.03 & \phn90.4/\phn7.5 & 0.07 & \phn6.7\\

$B$(4) & -0.52/0.07 & -0.10/0.07 & \phm{-}0.54/0.06 & \phn95.6/\phn4.0 & 0.06 & \phn4.6\\

$V$(4) & -0.53/0.05 & -0.07/0.05 & \phm{-}0.54/0.05 & \phn93.7/\phn2.7 & 0.05 & \phn5.2\\

$R$(4) & -0.50/0.07 & -0.06/0.03 & \phm{-}0.52/0.08 & \phn93.3/\phn1.2 & 0.06 & \phn5.9\\

$I$(4) & -0.48/0.08 & -0.10/0.06 & \phm{-}0.52/0.05 & \phn95.9/\phn4.0 & 0.14 & \phn9.4\\

\enddata
\end{deluxetable}

\begin{deluxetable}{lcccccccccc}
\tablecolumns{11}
\tablewidth{0pt}
\tablecaption{Statistical Summary}
\tablehead{
\colhead{Star} &
\colhead{$q\/$/$dq$} &
\colhead{\tt <$dq$\tt >} &
\colhead{$u\/$/$du$} &
\colhead{\tt <$du$\tt >} &
\colhead{$p\/$/$dp$} &
\colhead{\tt <$dp$\tt >} &
\colhead{$\theta\/$/$d\theta$} &
\colhead{\tt <$d\theta$\tt >} &
\colhead{\tt <$dpi$\tt >} &
\colhead{\tt <$d\theta i$\tt >} \\
\colhead{Filter} &
\colhead{(\%)} &
\colhead{(\%)} &
\colhead{(\%)} &
\colhead{(\%)} &
\colhead{(\%)} &
\colhead{(\%)} &
\colhead{($\arcdeg$)} & 
\colhead{($\arcdeg$)} &
\colhead{(\%)} &
\colhead{($\arcdeg$)}
}

\startdata

\sidehead{$\xi$ Per}

$U$ & -0.88/0.07 & 0.08 & -0.60/0.04 & 0.09 & \phm{-}1.07/0.07 & 0.09 & 107.2/\phn0.9 & \phn1.7 & 0.11 & \phn1.5\\

$B$ & -1.02/0.08 & 0.08 & -0.73/0.08 & 0.10 & \phm{-}1.26/0.08 & 0.09 & 107.8/\phn1.8 & \phn2.0 & 0.09 & \phn2.2\\

$V$ & -1.09/0.06 & 0.06 & -0.82/0.06 & 0.07 & \phm{-}1.37/0.08 & 0.06 & 108.5/\phn0.8 & \phn1.6 & 0.05 & \phn1.5\\

$R$ & -1.13/0.05 & 0.08 & -0.82/0.10 & 0.05 & \phm{-}1.40/0.09 & 0.07 & 108.0/\phn1.2 & \phn1.2 & 0.08 & \phn2.0\\

$I$ & -1.03/0.04 & 0.08 & -0.73/0.05 & 0.10 & \phm{-}1.26/0.04 & 0.05 & 107.6/\phn1.3 & \phn2.4 & 0.07 & \phn2.2\\

GAV & \nodata/0.06 & 0.07 & \nodata/0.07 & 0.08 & \nodata/0.07 & 0.07 & \nodata/\phn1.2 & \phn1.8 & 0.08 & \phn1.9\\

\sidehead{$\alpha$ Cam}

$U$ & \phm{-}0.18/0.03 & 0.10 & -1.52/0.21 & 0.12 & \phm{-}1.53/0.21 & 0.12 & 138.4/\phn0.0 & \phn2.7 & 0.11 & \phn1.8\\

$B$ & \phm{-}0.23/0.09 & 0.11 & -1.60/0.10 & 0.12 & \phm{-}1.62/0.11 & 0.11 & 139.0/\phn1.4 & \phn1.7 & 0.09 & \phn1.5\\

$V$ & \phm{-}0.13/0.11 & 0.08 & -1.58/0.21 & 0.08 & \phm{-}1.59/0.21 & 0.08 & 137.2/\phn1.7 & \phn1.5 & 0.06 & \phn0.9\\

$R$ & \phm{-}0.09/0.06 & 0.11 & -1.47/0.10 & 0.13 & \phm{-}1.48/0.10 & 0.12 & 136.7/\phn1.3 & \phn1.5 & 0.06 & \phn1.2\\

$I$ & \phm{-}0.06/0.03 & 0.08 & -1.23/0.15 & 0.08 & \phm{-}1.24/0.15 & 0.08 & 136.4/\phn0.8 & \phn2.1 & 0.06 & \phn2.0\\

GAV & \nodata/0.06 & 0.10 & \nodata/0.16 & 0.11 & \nodata/0.15 & 0.10 & \nodata/\phn1.1 & \phn1.9 & 0.08 & \phn1.5\\

\sidehead{$\lambda$ Ori}

$U$ & -0.23/0.02 & 0.02 & -0.14/0.05 & 0.08 & \phm{-}0.28/0.04 & 0.05 & 104.0/\phn3.1 & \phn6.0 & 0.04 & \phn5.4\\

$B$ & -0.23/0.05 & 0.07 & -0.17/0.05 & 0.08 & \phm{-}0.30/0.06 & 0.07 & 108.2/\phn5.8 & \phn8.2 & 0.07 & \phn8.8\\

$V$ & -0.25/0.02 & 0.04 & -0.14/0.03 & 0.05 & \phm{-}0.31/0.01 & 0.04 & 104.4/\phn3.2 & \phn4.7 & 0.03 & \phn6.8\\

$R$ & -0.25/0.03 & 0.05 & -0.11/0.02 & 0.05 & \phm{-}0.28/0.05 & 0.06 & 102.1/\phn0.1 & \phn4.9 & 0.03 & \phn5.3\\

$I$ & -0.25/0.03 & 0.05 & -0.06/0.01 & 0.08 & \phm{-}0.28/0.06 & 0.05 & \phn97.1/\phn0.6 & \phn7.9 & 0.05 & \phn7.1\\

GAV & \nodata/0.03 & 0.05 & \nodata/0.03 & 0.07 & \nodata/0.04 & 0.05 & \nodata/\phn2.6 & \phn6.3 & 0.04 & \phn6.7\\

\sidehead{$\zeta$ Ori}

$U$ & -0.19/0.06 & 0.06 & \phm{-}0.06/0.05 & 0.06 & \phm{-}0.23/0.07 & 0.05 & \phn77.3/\phn6.4 & \phn8.0 & 0.05 & \phn6.6\\

$B$ & -0.19/0.06 & 0.11 & \phm{-}0.07/0.07 & 0.06 & \phm{-}0.23/0.04 & 0.08 & \phn79.2/12.1 & 12.6 & 0.07 & 13.6\\

$V$ & -0.25/0.11 & 0.07 & \phm{-}0.10/0.03 & 0.05 & \phm{-}0.29/0.09 & 0.06 & \phn79.8/\phn5.6 & \phn7.2 & 0.04 & 10.2\\

$R$ & -0.20/0.01 & 0.06 & \phm{-}0.09/0.01 & 0.06 & \phm{-}0.23/0.02 & 0.08 & \phn80.4/\phn1.1 & \phn7.2 & 0.05 & \phn6.9\\

$I$ & -0.17/0.08 & 0.09 & \phm{-}0.11/0.11 & 0.05 & \phm{-}0.25/0.04 & 0.08 & \phn73.3/19.2 & \phn8.2 & 0.04 & 10.2\\

GAV & \nodata/0.06 & 0.08 & \nodata/0.05 & 0.06 & \nodata/0.05 & 0.07 & \nodata/\phn8.9 & \phn8.6 & 0.05 & \phn9.5\\

\sidehead{$\zeta$ Oph}

$U$ & -0.24/0.05 & 0.05 & -1.05/0.07 & 0.06 & \phm{-}1.08/0.07 & 0.06 & 128.6/\phn1.2 & \phn0.7 & 0.05 & \phn1.1\\

$B$ & -0.34/0.04 & 0.04 & -1.27/0.07 & 0.04 & \phm{-}1.31/0.07 & 0.04 & 127.6/\phn0.8 & \phn0.8 & 0.04 & \phn0.9\\

$V$ & -0.46/0.10 & 0.03 & -1.39/0.05 & 0.02 & \phm{-}1.47/0.03 & 0.02 & 125.9/\phn2.0 & \phn0.7 & 0.04 & \phn0.7\\

$R$ & -0.45/0.11 & 0.04 & -1.39/0.05 & 0.04 & \phm{-}1.46/0.04 & 0.05 & 126.0/\phn2.2 & \phn0.6 & 0.04 & \phn0.7\\

$I$ & -0.45/0.09 & 0.03 & -1.29/0.12 & 0.04 & \phm{-}1.37/0.14 & 0.04 & 125.5/\phn1.3 & \phn0.8 & 0.04 & \phn1.1\\

GAV & \nodata/0.08 & 0.04 & \nodata/0.07 & 0.04 & \nodata/0.07 & 0.04 & \nodata/\phn1.5 & \phn0.7 & 0.04 & \phn0.9\\

\sidehead{68 Cyg}

$U$ & -0.28/0.05 & 0.04 & \phm{-}0.47/0.06 & 0.06 & \phm{-}0.56/0.05 & 0.05 & \phn60.3/\phn3.4 & \phn2.6 & 0.09 & \phn4.6\\

$B$ & -0.29/0.10 & 0.06 & \phm{-}0.44/0.06 & 0.07 & \phm{-}0.54/0.10 & 0.06 & \phn61.1/\phn3.9 & \phn4.7 & 0.07 & \phn4.5\\

$V$ & -0.34/0.05 & 0.04 & \phm{-}0.51/0.02 & 0.04 & \phm{-}0.62/0.03 & 0.03 & \phn61.7/\phn2.3 & \phn2.0 & 0.05 & \phn1.8\\

$R$ & -0.37/0.03 & 0.04 & \phm{-}0.48/0.05 & 0.04 & \phm{-}0.61/0.04 & 0.04 & \phn63.7/\phn2.2 & \phn2.1 & 0.04 & \phn2.0\\

$I$ & -0.36/0.06 & 0.06 & \phm{-}0.45/0.05 & 0.07 & \phm{-}0.58/0.02 & 0.08 & \phn64.0/\phn3.5 & \phn2.2 & 0.09 & \phn4.9\\

GAV & \nodata/0.06 & 0.05 & \nodata/0.05 & 0.06 & \nodata/0.05 & 0.05 & \nodata/\phn3.1 & \phn2.7 & 0.07 & \phn3.5\\

\sidehead{19 Cep}

$U$ & -0.88/0.00 & 0.12 & \phm{-}0.67/0.00 & 0.10 & \phm{-}1.12/0.00 & 0.15 & \phn71.4/\phn0.0 & \phn0.2 & 0.12 & \phn2.5\\

$B$ & -0.92/0.12 & 0.10 & \phm{-}0.68/0.09 & 0.10 & \phm{-}1.15/0.10 & 0.10 & \phn71.7/\phn2.6 & \phn2.5 & 0.09 & \phn2.6\\

$V$ & -0.89/0.01 & 0.06 & \phm{-}0.67/0.15 & 0.08 & \phm{-}1.12/0.08 & 0.05 & \phn71.6/\phn3.3 & \phn2.0 & 0.05 & \phn0.9\\

$R$ & -0.83/0.00 & 0.04 & \phm{-}0.55/0.00 & 0.06 & \phm{-}1.01/0.00 & 0.04 & \phn73.4/\phn0.0 & \phn1.7 & 0.04 & \phn4.2\\

$I$ & -0.73/0.00 & 0.10 & \phm{-}0.45/0.00 & 0.04 & \phm{-}0.86/0.00 & 0.10 & \phn74.2/\phn0.0 & \phn1.9 & 0.15 & \phn3.8\\

GAV & \nodata/0.03 & 0.08 & \nodata/0.05 & 0.07 & \nodata/0.04 & 0.09 & \nodata/\phn1.2 & \phn1.7 & 0.09 & \phn2.8\\

\sidehead{$\lambda$ Cep}

$U$ & -0.51/0.10 & 0.09 & \phm{-}1.00/0.02 & 0.09 & \phm{-}1.13/0.06 & 0.09 & \phn58.4/\phn2.1 & \phn2.3 & 0.10 & \phn2.3\\

$B$ & -0.51/0.04 & 0.08 & \phm{-}1.09/0.04 & 0.07 & \phm{-}1.20/0.02 & 0.08 & \phn57.4/\phn1.2 & \phn1.8 & 0.06 & \phn1.6\\

$V$ & -0.48/0.05 & 0.04 & \phm{-}1.09/0.04 & 0.07 & \phm{-}1.19/0.03 & 0.07 & \phn57.1/\phn1.3 & \phn1.3 & 0.04 & \phn1.0\\

$R$ & -0.53/0.10 & 0.06 & \phm{-}1.10/0.06 & 0.04 & \phm{-}1.22/0.10 & 0.03 & \phn57.7/\phn1.5 & \phn1.5 & 0.05 & \phn1.3\\

$I$ & -0.41/0.08 & 0.09 & \phm{-}0.95/0.10 & 0.09 & \phm{-}1.04/0.11 & 0.09 & \phn56.5/\phn1.2 & \phn2.7 & 0.09 & \phn2.5\\

GAV & \nodata/0.07 & 0.07 & \nodata/0.05 & 0.07 & \nodata/0.06 & 0.07 & \nodata/\phn1.5 & \phn1.9 & 0.07 & \phn1.7\\

\sidehead{10 Lac}

$U$ & -0.42/0.04 & 0.05 & -0.04/0.05 & 0.06 & \phm{-}0.44/0.04 & 0.05 & \phn87.6/\phn4.0 & \phn4.8 & 0.06 & \phn4.9\\

$B$ & -0.50/0.02 & 0.08 & -0.07/0.06 & 0.08 & \phm{-}0.52/0.03 & 0.07 & \phn89.9/\phn5.8 & \phn4.9 & 0.07 & \phn4.5\\

$V$ & -0.52/0.01 & 0.04 & -0.05/0.02 & 0.04 & \phm{-}0.53/0.01 & 0.04 & \phn90.8/\phn4.1 & \phn2.2 & 0.05 & \phn3.2\\

$R$ & -0.51/0.01 & 0.06 & -0.04/0.04 & 0.02 & \phm{-}0.52/0.00 & 0.07 & \phn91.4/\phn2.6 & \phn1.2 & 0.05 & \phn4.0\\

$I$ & -0.47/0.01 & 0.06 & -0.06/0.05 & 0.05 & \phm{-}0.50/0.04 & 0.05 & \phn92.1/\phn5.4 & \phn3.0 & 0.09 & \phn5.9\\

GAV & \nodata/0.02 & 0.06 & \nodata/0.04 & 0.05 & \nodata/0.02 & 0.06 & \nodata/\phn4.4 & \phn3.2 & 0.06 & \phn4.5\\

\enddata
\end{deluxetable}

\end{document}